\documentclass[twocolumn]{jpsj2} 

\title{Geometrical Quadrupolar Frustration in DyB$_4$}

\author{
Ryuta~\textsc{Watanuki} \thanks{E-mail address: ryu-wat@issp.u-tokyo.ac.jp},
Gou~\textsc{Sato}$^{1}$, 
Kazuya~\textsc{Suzuki}$^{1}$, 
Masaki~\textsc{Ishihara}$^{2}$, \\
Tatsuya~\textsc{Yanagisawa}$^{2}$, 
Yuichi~\textsc{Nemoto}$^{2}$ and 
Terutaka~\textsc{Goto}$^{2}$
}

\inst{
Institute for Solid State Physics, University of Tokyo, Kashiwa, Chiba 277--8581, Japan\\
$^{1}$ Faculty of Engineering, Yokohama National University, Yokohama 240--8501, Japan\\
$^{2}$ Graduate School of Science and Technology, Niigata University, Niigata 950--2181, Japan
}

\recdate{\today}

\abst{
Physical properties of DyB$_4$ have been studied by magnetization, specific heat, and ultrasonic measurements. The magnetic entropy change and ultrasonic properties in intermediate phase II indicate that the degeneracy of internal degrees of freedom is not fully lifted in spite of the formation of magnetic order. The ultrasonic attenuation and huge softening of $C_{44}$ in phase II suggest the existence of electric-quadrupolar (orbital) fluctuations of 4$f$-electrons. These unusual properties originate from a geometrical quadrupolar frustration.
}

\kword{DyB$_4$, quadrupole order, Shastry-Sutherland lattice, geometrical quadrupolar frustration, quadrupolar fluctuation}

\begin{document}
\maketitle

A geometrical frustration, that is, a geometrical prohibition against satisfying the lowest-energy state of each bipartite bond, leads to various unusual physical properties, that often cause an extensive degeneracy in the ground state of a system and prevent ordering. A well-known example is the spin frustration in triangular antiferromagnets, such as CsCoCl$_3$ and CsNiCl$_3$, that causes various magnetic-ordering processes. \cite{CsCoCl3, CsNiCl3-1, CsNiCl3-2, CsNiCl3-3, CsNiCl3-4}$^)$ A renewed interest in geometrical frustration systems has resulted from the discoveries of new magnetic properties in certain materials. Two examples of these materials are the pyrochlore oxides Dy$_2$Ti$_2$O$_7$ and Ho$_2$Ti$_2$O$_7$ considered to be model systems of spin ice materials. \cite{Spin-ice-1,Spin-ice-2}$^)$ A spin ice model shows a strong geometrical frustration arising from the competition between a ferromagnetic nearest-neighbor interaction and an axial single-ion anisotropy. The spins in this model are disordered with local configurations obeying the ice rule in every tetrahedron, and the ground state is macroscopically degenerated with a residual entropy. In fact, Dy$_2$Ti$_2$O$_7$ shows a residual entropy of $\sim 1.68$~J/mol~K, which is numerically in agreement with Pauling's entropy for water ice, $R (1/2)\ln (3/2)$. \cite{Spin-ice-3}$^)$ Another new noteworthy discovery is the formation of unusual short-range order named the magnetic random layer (MRL) lattice in RB$_2$C (R = Dy, Ho), \cite{RB2C-PRL}$^)$ that results in an unusual neutron scattering in the low-wave-number region. The unconventional MRL state with large magnetic fluctuations originates from the frustrated nature of the Shastry-Sutherland lattice (SSL) consisting of the rare-earth atoms. \cite{RB2C-PRB}$^)$

\begin{figure}[tb]
\begin{center}
\includegraphics[width=65mm]{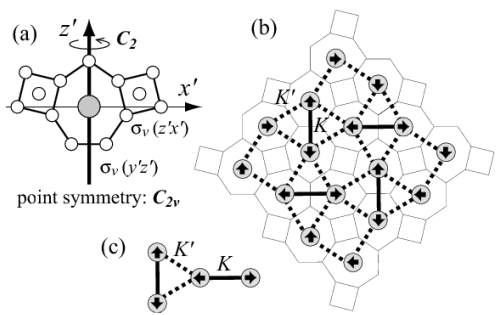}
\end{center}
\caption{(a) Local structure of Dy$^{3+}$ ion in DyB$_4$. The Dy$^{3+}$ ion is indicated by the gray circle and the boron atoms by the open circles. (b) Lattice structure of the Dy$^{3+}$ ions of DyB$_4$. The bold solid lines are the inter square-lattice bonds $K$, and the bold broken lines are the intra square-lattice bonds $K'$. The thin solid lines represent the boron frame. The twofold axes of the local symmetry at the Dy sites are represented by the arrows. (c) Interactions of the elementary unit of the SSL.} 
\label{fig:DyB4_model}
\end{figure}

In this letter, we focus on a frustration of the orbital (quadrupole) degrees of freedom for 4$f$ electron system, and propose that DyB$_4$ is the first example of \textit{a geometrically quadrupolar (orbital) frustrated system}. DyB$_4$ crystallizes in the tetragonal ThB$_4$-type structure belonging to space group $D^5_{4h}-P4/mbm$. \cite{ThB4-structure, LaB4-structure-X, Er-DyB4-structure-N}$^)$ The boron atoms form a continuous three-dimensional network. The Dy atoms are located below and above the centers of the seven-membered boron rings in densely packed planes of boron atoms. All Dy atoms occupy the equivalent sites having orthorhombic symmetry $C_{2v}$ (Fig.\ \ref{fig:DyB4_model} (a)). The two-dimensional layer of the Dy atoms is formed by fused equilateral triangles and squares. It can be considered that this arrangement of the Dy atoms is the SSL. Previous studies of the magnetic properties of DyB$_4$ have shown that it orders antiferromagnetically. \cite{rb4-mag-bushow,Etourneau}$^)$ Two second-order magnetic phase transitions have been discovered at 20.4~K and 12.8~K. \cite{RB4-Fisk-ru-chi}$^)$ A neutron diffraction study of Dy$^{11}$B$_4$ was performed at 4.2~K. \cite{ErB4-DyB4-neut}$^)$ The deduced magnetic structure is a simple collinear antiferromagnetic one with the sequence $(++--)$ for the four Dy atoms in the tetragonal unit cell. The magnetic moments are oriented parallel to the $[001]$-axis. However, it seems that the data is insufficient for determining the detailed magnetic structure of DyB$_4$, owing to the strong absorption of thermal neutrons by dysprosium. There are very few reports on the physical properties of DyB$_4$ and studies from the point of view of the frustration and the role of quadrupolar interactions are lacking. Here we investigate the details of the physical properties of DyB$_4$ by means of magnetic susceptibility, specific heat, and ultrasonic measurements.

Single crystals of DyB$_4$ were prepared by the Czochralski method in argon atmosphere using a tetra-arc furnace. The magnetization of DyB$_4$ was measured using a SQUID magnetometer in the temperature range of 1.8--300~K. The magnetic susceptibility measurements were performed under zero-field-cooled (ZFC) and field-cooled (FC) conditions. Specific heat measurement was carried out by the heat-relaxation method down to 1.9~K. The elastic properties of DyB$_4$ were investigated by means of ultrasonic measurements. Ultrasonic measurement is an effective technique for examining the quadrupolar effect in an $f$-electron system with an orbitally degenerate ground state, because the quadrupole of the 4$f$ electron couples to the elastic strain. The elastic constant $C_{\Gamma}$ of the sample was obtained by $C_{\Gamma} = \rho v^2$, where $v$ and $\rho$ represent the sound velocity and the mass density, respectively. $v$ was measured by a phase comparator using double balanced mixers and $\rho= 6.783 \times 10^3 $~kg/m$^3$ was calculated using the lattice constants $a= 7.0960(4)$~\AA~and $c= 4.0128(2)$~\AA~of DyB$_4$ at 25~K. \cite{Neutron_DyB4}$^)$

\begin{figure}[tb]
\begin{center}
\includegraphics[width=63mm]{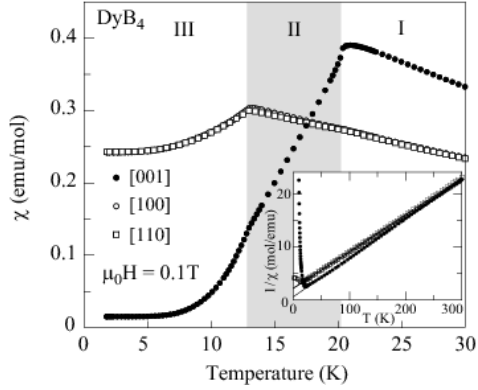}
\end{center}
\caption{Magnetic susceptibility $\chi$ of DyB$_4$. The inset shows the Curie-Weiss fit to the inverse susceptibility.}
\label{fig:DyB4_MT}
\end{figure}

Figure~\ref{fig:DyB4_MT} shows the magnetic susceptibility $\chi$ for a single crystal of DyB$_4$. No difference between the FC and ZFC susceptibilities in any direction is observed throughout the whole temperature range; therefore, only ZFC data are shown for clarification. The coincidence of the ZFC and FC data suggests that no spontaneous magnetization is present in DyB$_4$. The susceptibility for $\mbox{\boldmath $H$} \parallel [001]$ shows a maximum at about 21.0~K, and a small discontinuous drop occurs at 20.3~K. In addition, a small shoulder is observed at 13.0~K. On the other hand, susceptibilities for $\mbox{\boldmath $H$} \parallel [100]$ and $\mbox{\boldmath $H$} \parallel [110]$ show no anomalies around 21~K, but exhibit cusplike maxima at 13.0~K. These results suggest that the $z$ and $xy$ components of the magnetic moments order independently at $T_{\rm C1} \sim 20.3$~K and $T_{\rm C2} \sim 13.0$~K, respectively. The results may be divided into three temperature regions, which are hereafter called phase I ($T>T_{\rm C1}$), II ($T_{\rm C2} <T< T_{\rm C1}$) and III ($T<T_{\rm C2}$) in the sequence of decreasing temperature. The magnetic structure in phase II is a partially dynamical one, that is, the components of the magnetic moments parallel to the $[001]$-axis are ordered while the perpendicular components remain disordered; then, below $T_{\rm C2}$, all components are ordered.

The inverse molar susceptibilities $\chi^{-1}$ of DyB$_4$ (the inset of Fig.\ \ref{fig:DyB4_MT}) show linear temperature dependences above 250~K for $\mbox{\boldmath $H$} \parallel [001]$ and 35~K for $\mbox{\boldmath $H$} \parallel [100]$ and $\mbox{\boldmath $H$} \parallel [110]$. The effective paramagnetic moments $\mu_{\rm eff}$ and the Weiss temperatures $\theta_{\rm p}$ resulting from the Curie-Weiss fitting are as follows: $\mu_{\rm eff}^{\parallel [001]} = 10.44$~$\mu_{\rm B}$ and $\theta_{\rm p}^{\parallel [001]} = -8.46$~K; $\mu_{\rm eff}^{\parallel [100]} = 10.90$~$\mu_{\rm B}$ and $\theta_{\rm p}^{\parallel [100]} = -33.40$~K; $\mu_{\rm eff}^{\parallel [110]} = 10.75$~$\mu_{\rm B}$ and $\theta_{\rm p}^{\parallel [110]} = -31.42$~K. Since the effective paramagnetic moments are close to that expected for a free Dy$^{3+}$ ion, i.e., 10.63~$\mu_{\rm B}$, we conclude that the 4$f$ electrons in DyB$_4$ are well localized. The antiferromagnetic $\theta_{\rm p}$ are comparable to $T_{\rm C1}$. These indicate that the spin frustration is small in DyB$_4$. 

\begin{figure}[tb]
\begin{center}
\includegraphics[width=68mm]{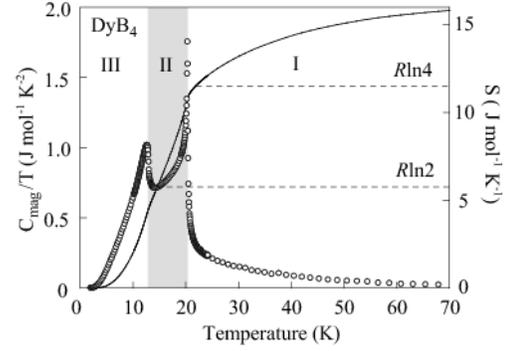}
\end{center}
\caption{Magnetic specific heat divided by temperature $C_{\rm mag}/T$ (left axis) and corresponding entropy $S$ (right axis) vs temperature for DyB$_4$. The horizontal broken lines indicate the values of entropy for one and two Kramers doublets.}
\label{fig:DyB4_SH}
\end{figure}

\begin{figure}[tb]
\begin{center}
\includegraphics[width=65mm]{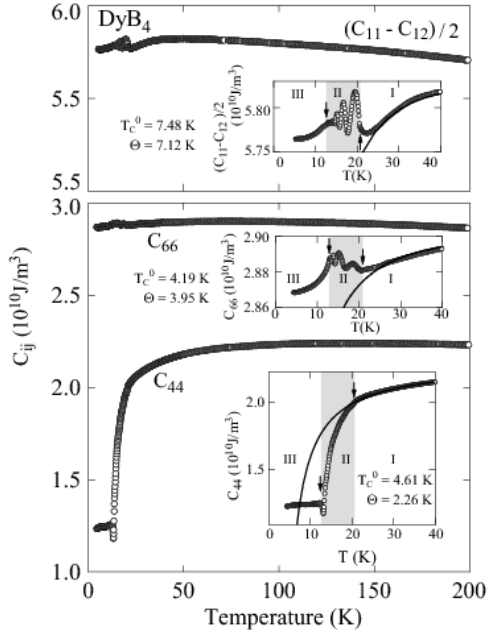}
\end{center}
\caption{Temperature dependence of the elastic constants of DyB$_4$. Transverse modes $(C_{11}-C_{12})/2$, $C_{44}$, and $C_{66}$ were measured at frequencies of 8~MHz. The insets show the low-temperature data of the transverse modes. The solid lines are fits to the data using $C_{\Gamma}(T)=C_{\Gamma}^{0}(T-T_{\rm C}^0)/(T-\Theta)$.}
\label{fig:DyB4_elastic_Cij}
\end{figure}

The specific heat of the single-crystalline sample of DyB$_4$ shows two $\lambda$-like anomalies that indicate exact transition temperatures of $T_{\rm C1}=20.3$~K and $T_{\rm C2}=12.7$~K. These transition temperatures are consistent with those reported previously. \cite{RB4-Fisk-ru-chi}$^)$ Since the peak of the specific heat at $T_{\rm C1}$ is very sharp, and moreover, $\chi$ for $\mbox{\boldmath $H$} \parallel [001]$ shows the discontinuity at $T_{\rm C1}$, the phase transition at $T_{\rm C1}$ seems to be a first-order one. In our present study, however, we cannot conclude whether this phase transition is a first-order or a second-order transition because the discontinuity of $\chi$ at $T_{\rm C1}$ is very small and the elastic constants are continuous at $T_{\rm C1}$ (Fig. \ref{fig:DyB4_elastic_Cij}). Only the $z$ components of the magnetic moments order at $T_{\rm C1}$; that is, the Z2-symmetry of the system breaks in phase II. Therefore, the phase transition at $T_{\rm C1}$ may be a second-order one and may belong to the three-dimensional Ising universality class in which the peak of the specific heat at a phase transition is very sharp. The phase transition at $T_{\rm C2}$ is a typical second-order one. The temperature dependence of the magnetic specific heat divided by temperature $C_{\rm mag}/T$ is shown in Fig. \ref{fig:DyB4_SH}. The estimated magnetic contribution $C_{\rm mag}$ to the specific heat of DyB$_4$ is obtained by subtracting the phononic part from the total specific heat of DyB$_4$; the phonon contribution is obtained from a measurement of the isostructural and nonmagnetic compound LuB$_4$. We obtain the temperature dependence of the magnetic entropy per Dy$^{3+}$ ion $S$ by numerically integrating the data of $C_{\rm mag}/T$ vs $T$. The magnetic entropy of approximately $R \ln 2$ and $R \ln 4$ is released below $T_{\rm C2}$ and $T_{\rm C1}$, respectively. The CEF ground-state multiplet $J=15/2$ of Dy$^{3+}$ splits into eight $E_{1/2}$ ($\Gamma_5$) doublets in the CEF potential of the site with $C_{2v}$ symmetry in DyB$_4$. The result indicates that the CEF ground state of DyB$_4$ is a pseudo-quartet consisting of two Kramers doublets with $E_{1/2}$ ($\Gamma_5$) symmetry. It is noted that the degeneracy due to the internal degrees of freedom is not fully lifted in spite of the formation of magnetic order in phase II. A similar entropy change is shown in the quadrupolar compound DyB$_2$C$_2$. \cite{DyB2C2}$^)$ In this case, the twofold Kramers degeneracy is conserved since time-reversal symmetry is not broken by the antiferroquadrupolar ordering. On the other hand, in a conventional magnet, degeneracy is lifted below the magnetic transition temperature since Zeeman splitting caused by an internal magnetic field is different in each Kramers doublet. Hence, phase II of DyB$_4$ is not the conventional magnetically ordered state. 

Elastic constants of DyB$_4$ (Fig. \ref{fig:DyB4_elastic_Cij}) show clear anomalies associated with the successive transitions at $T_{\rm C1}$ and $T_{\rm C2}$. The transverse mode $C_{44}$ associated with elastic strain $\varepsilon_{yz}$ or $\varepsilon_{zx}$ exhibits a large softening of 10.4\% below 120~K down to $T_{\rm C1}$. On the other hand, the transverse mode $(C_{11}-C_{12})/2$ and $C_{66}$ show small softening of 0.8\% and 0.6\% between 70~K and $T_{\rm C1}$, respectively. The largest softening of $C_{44}$ in phase I indicates that the dominant components of quadrupole moments in DyB$_4$ are $O_{yz} = J_y J_z+J_z J_y$ and $O_{zx} = J_z J_x+J_x J_z$ which couple with the elastic strains $\varepsilon_{yz}$ and $\varepsilon_{zx}$, respectively. Furthermore, in phase I, the elastic softenings are described by the equation $C_{\Gamma}(T)=C_{\Gamma}^{0}(T-T_{\rm C}^0)/(T-\Theta)$, which is deduced from the Curie term of the quadrupole-strain susceptibility. Here, $\Theta$ is proportional to the average quadrupole-quadrupole interaction between the Dy$^{3+}$ ions. $T_{\rm C}^0 = \Theta + E_{\rm JT}$, where $E_{\rm JT}$ is the Jahn-Teller (4$f$ electron--lattice) coupling. \cite{quadrupole-strain susceptibility}$^)$ The characteristic temperatures obtained are as follows: $\Theta = 2.26$~K and $T_{\rm C}^0 = 4.61$~K for $C_{44}$; $\Theta = 3.95$~K and $T_{\rm C}^0 = 4.19$~K for $C_{66}$; $\Theta = 7.12$~K and $T_{\rm C}^0 = 7.48$~K for $(C_{11}-C_{12})/2$. The positive values of $\Theta$ indicate that the predominant quadrupolar interactions in DyB$_4$ are ferro-type. $C_{44}$ shows a kink at $T_{\rm C1}$ and decreases abruptly with decreasing temperature from $T_{\rm C1}$ to $T_{\rm C2}$. It should be noted that $C_{44}$ shows a marked softening of 40\% and its echo signal shrinks considerably throughout the whole temperature range of magnetically ordered phase II. With decreasing temperature, the elastic constants change their behavior at $T_{\rm C2}$ from softening to hardening, and then almost level off towards 0~K. The echo signal is recovered and stabilized below $T_{\rm C2}$. 

The huge elastic softenings and the ultrasonic absorption of $C_{44}$ in phase II are different from the critical behaviors observed in the vicinity of a structural phase transition temperature induced by either ferroquadrupolar ordering (e.g., DyB$_6$, HoB$_6$ \cite{DyB6-HoB6}$^)$) or the cooperative Jahn-Teller effect (e.g., DyVO$_4$ \cite{DyVO4}$^)$). On the other hand, it has been known that large elastic softenings and ultrasonic absorptions in an ordered state arise from magnetic and structural domain-wall stress effects. \cite{Luthi}$^)$ However, in DyB$_4$, no symmetry reductions of the crystal and magnetic structure at the transition from phase II to phase III were observed in the neutron diffraction study. \cite{Neutron_DyB4}$^)$ Therefore, the elastic instability observed in only phase II of DyB$_4$ also cannot be explained by domain-wall stress effects. Hence, we conclude that the attenuation of the echo signal of elastic constants in phase II of DyB$_4$ is caused by quadrupolar fluctuations and that the orbital degeneracy is still conserved in phase II. This is consistent with the results of the magnetic entropy change. On the contrary, the elastic behaviors in phase III of DyB$_4$ indicate that the degeneracy of the orbital degrees of freedom is lifted and the quadrupolar fluctuations disappear. Consequently, quadrupolar ordering and antiferromagnetic ordering may coexist below $T_{\rm C2}$.

Hereafter, we focus on the quadrupolar instability in phase II of DyB$_4$, and we hypothesize that the quadrupolar fluctuations in DyB$_4$ can be caused by a frustration in the 4$f$ orbital (quadrupolar) system of DyB$_4$. In the following, we consider two possible origins of the quadrupolar frustration in DyB$_4$. The first is based on a competition between the quadrupole-quadrupole interactions on the SSL formed by Dy$^{3+}$ ions in the $xy$-layer. An elementary unit model of the SSL for the nearest-neighbor interactions is shown in Fig.\ \ref{fig:DyB4_model} (c). If the coupling $K$ is of antiferro-type, the quadrupolar system is frustrated regardless of the sign of the coupling $K'$, as in a spin system. The second is that the uniaxial anisotropy effects of the 4$f$ orbitals compete with the ferroquadrupolar interactions. The single ion anisotropy of the 4$f$ orbital probably reflects the local environment of the rare-earth site through 4$f$ electron-lattice coupling and/or CEF effects. The local symmetry $C_{2v}$ of the Dy site is characterized by a twofold axis ($z'$-axis in Fig.\ \ref{fig:DyB4_model} (a)) lying in the $xy$-plane. Because the site symmetry of the Dy site is uniaxial, it is reasonable to assume that the quadrupolar anisotropy is also uniaxial. The most important point to note is the geometrical arrangement of the twofold axis at the Dy site. In Figs. \ref{fig:DyB4_model} (b) and \ref{fig:DyB4_model} (c), the twofold axes at the Dy sites are represented by arrows. The two twofold axes of a Dy-pair coupled by the $K$-bond are collinear and the directions are antiparallel to each other. In contrast, the arrows between the pairs are orthogonal to each other. If only the single-ion anisotropy of Dy$^{3+}$ is considered, 4$f$ orbitals on the neighboring sites are orthogonal to each other. Therefore, the single-ion anisotropy effects of the 4$f$ orbital in DyB$_4$ compete with the ferroquadrupolar interactions. In addition, the present results that $\Theta$ are comparable to $E_{\rm JT}$ suggest that this competition is strong in DyB$_4$. Accordingly, it can be expected that the quadrupolar system of DyB$_4$ is geometrically frustrated and shows a quadrupolar instability. It can be said that this quadrupolar frustration model is similar to the spin-ice model which shows that strong geometrical frustration can arise in presence of the dominant ferromagnetic interactions and local Ising anisotropy. \cite{Spin-ice-1,Spin-ice-2}$^)$  We obtained the result that predominant quadrupolar interactions in DyB$_4$ are of ferro-type. However, these results can be obtained if the coupling $K'$ and the interlayer interactions are ferro-type because $\theta$ is only a mean value of the interaction. At present, the information is insufficient to conclude which origin of the frustration is more credible. 

In the following, we compare DyB$_4$ and RB$_2$C (R = Dy and Ho) in which the rare-earth ions also form the SSL. DyB$_2$C and HoB$_2$C show a three-dimensional magnetic order but MRL states still remain below $T_{\rm C}$. \cite{RB2C-PRB}$^)$ The magnetic ordering temperatures of RB$_2$C (R = Dy, Ho, and Er) do not follow the de Gennes relation. This fact suggests that the ordering temperatures of DyB$_2$C and HoB$_2$C are suppressed. The unconventional fluctuating ground states and the suppressions of the ordering temperatures of DyB$_2$C and HoB$_2$C are due to spin frustration effects. In DyB$_4$, however, the large magnetic fluctuations such as those observed in DyB$_2$C and HoB$_2$C were not recognized from the results of powder neutron diffraction measurement \cite{Neutron_DyB4}$^)$; this is consistent with the results of the inverse magnetic susceptibilities. Consequently, it is thought that the influence of spin frustration on the magnetic transition temperature is very small in this compound. It seems that the quadrupolar system of DyB$_4$ is strongly influenced by a geometrical quadrupolar frustration and that the quadrupolar ordering is suppressed. 

Most of the compounds with spin frustration show long-range order in the lowest-temperature phase. However, a spin frustration system increases the degree of order through successive phase transitions accompanied by high-entropy intermediate states with spin fluctuation, because the energy gain of the spin-spin correlation is suppressed by the spin frustration. For example, a partial disorder and an order of limited components of spins are generally observed in triangular lattice antiferromagnets such as CsCoCl$_3$ and CsNiCl$_3$. \cite{CsCoCl3,CsNiCl3-1,CsNiCl3-2,CsNiCl3-3,CsNiCl3-4}$^)$ CsNiCl$_3$ exhibits successive phase transitions similar to those of DyB$_4$, where the parallel and perpendicular components to the $c$-axis of Ni$^{2+}$ moments order at $T_{\rm N1}=4.8$~K and $T_{\rm N2}=4.4$~K, respectively. In this material, the large difference between the Weiss temperature $\theta_{\rm p}=-69$~K and $T_{\rm N1}$ is attributed to the spin frustration. \cite{CsNiCl3-4}$^)$ On the other hand, in DyB$_4$, the successive phase transitions characterized by a magnetic ordering of the $z$ and $xy$ components at independent temperatures are not caused by spin frustration. The quadrupole moments and the $xy$ components of the magnetic moments order at $T_{\rm C2}$ simultaneously. The quadrupolar interactions most likely play a dominant role in the phase transition at $T_{\rm C2}$, and the ordering of the $xy$ components of the magnetic moments coupled with the quadrupole ones can be induced by the ordering of the quadrupole moments. At present, however, it is unclear why the phase transition at $T_{\rm C2}$ is not a first-order but a second-order transition. Accordingly, we conclude that the successive phase transitions in DyB$_4$ are caused by the coexistence of quadrupolar frustration and the quadrupolar entropy effect. 

In summary, we observed successive phase transition at $T_{C1}=20.3$~K and $T_{C2}=12.7$~K in DyB$_4$. The successive phase transitions are characterized by a magnetic ordering of the $z$ and $xy$ components at independent temperatures. The magnetic entropy change indicates that the degeneracy of the internal degrees of freedom in phase II is still conserved in spite of the formation of magnetic order. The huge softening of $C_{44}$ and the attenuation in phase II suggest the existence of quadrupolar fluctuations. These unusual properties originate from a geometrical quadrupolar frustration.

We are grateful to T.~Sakakibara, J.~Custers, and N.~Todoroki for their helpful discussions.


\begin{thebibliography}{99}
\bibitem{Neutron_DyB4}
Recently, we have performed a detailed powder neutron diffraction study of DyB$_4$. The results will be published elsewhere.
\end{thebibliography}
\end{document}